\documentclass[journal,10pt]{IEEEtran} 
\usepackage[figurename=Fig.]{caption}
\usepackage{tikz}
\usetikzlibrary{calc, positioning, shapes.geometric, decorations.pathreplacing}
\usepackage{array}
\usepackage{verbatim}
\usetikzlibrary{calc, shapes, arrows, positioning}
\usepackage{authblk}
\usepackage{blindtext}
\usepackage{ifpdf}
\usepackage{graphicx}
\usepackage{tabulary}
\usepackage{amsmath,amsthm,amssymb}
\usepackage{amsmath}
\usepackage{kantlipsum}
\allowdisplaybreaks
\usepackage{cleveref}
\usepackage{lipsum}%
\usepackage{optidef}
\usepackage[english]{babel} 
\theoremstyle{plain}

\crefname{theorem}{Theorem}{theorem}
\crefname{lemma}{Lemma}{Lemmas}

\usepackage{cite}
\usepackage{dsfont}
\usepackage[justification=centering]{caption}
\usepackage{color}
\usepackage{algorithm}
\usepackage{algorithmicx, algpseudocode}
\usepackage{etoolbox}
\usepackage{subcaption}
\usepackage{balance}
\usepackage{empheq,etoolbox}
\usepackage[utf8]{inputenc}
\usepackage{multirow}
\usepackage{float}

\usepackage{amssymb}% http://ctan.org/pkg/amssymb
\usepackage{pifont}% http://ctan.org/pkg/pifont
\usepackage[hmargin=1.27cm,top=1.3cm,bottom=1.3cm]{geometry}

\usepackage{babel}
\floatstyle{plaintop}
\restylefloat{table}
\patchcmd{\subequations}
\usepackage{lipsum} % For some dummy text
\allowdisplaybreaks

\tikzset{brace/.style={decorate, decoration={brace}},
 brace mirrored/.style={decorate, decoration={brace,mirror}},
}

\newcounter{brace}
\newcounter{arrow}
\begin{document}
 \captionsetup[figure]{name={Fig.},labelsep=period}

%\title{Low-Complexity and Robust Resource Allocation for RIS-Empowered Systems}
\title{Resource Allocation for Pinching-Antenna Systems: State-of-the-Art, Key Techniques and Open Issues}

\bstctlcite{IEEEexample:BSTcontrol}
%\author{Ming Zeng, Ji Wang, Octavia A. Dobre, Zhiguo Ding, George K. Karagiannidis, Robert Schober, and H. Vincent Poor
\author{Ming Zeng, Ji Wang, Octavia A. Dobre, \textit{Fellow, IEEE}, Zhiguo Ding, \textit{Fellow, IEEE}, George K. Karagiannidis, \textit{Fellow, IEEE}, Robert Schober, \textit{Fellow, IEEE}, and H. Vincent Poor, \textit{Life Fellow, IEEE}
    \thanks{M. Zeng is with the Department of Electric and Computer Engineering, Laval University, Quebec City, Canada (email: ming.zeng@gel.ulaval.ca).}
    
    \thanks{J. Wang is with the Department of Electronics
and Information Engineering, Central China Normal University, Wuhan, China (e-mail: jiwang@ccnu.edu.cn).}

\thanks{O. A. Dobre is with the Faculty of Engineering and Applied Science, Memorial University, St. John’s, Canada (e-mail: odobre@mun.ca).}

\thanks{Z. Ding is with The University of Manchester, Manchester, UK. (e-mail: zhiguo.ding@manchester.ac.uk).}  

\thanks{G. K. Karagiannidis is with the Department of Electrical and Computer Engineering, Aristotle University of Thessaloniki, Thessaloniki, Greece (e-mail: geokarag@auth.gr).}

\thanks{R. Schober is with the Institute for Digital Communications, FriedrichAlexander-University Erlangen-Nürnberg (FAU), 91054 Erlangen, Germany (e-mail: robert.schober@fau.de).}

\thanks{H. V. Poor is with the Department of Electrical and Computer Engineering, Princeton University, Princeton, NJ 08544 USA (e-mail: poor@princeton.edu).}
}
\maketitle

\begin{abstract}
Pinching antennas have emerged as a promising technology for reconfiguring wireless propagation environments, particularly in high-frequency communication systems operating in the millimeter-wave and terahertz bands. By enabling dynamic activation at arbitrary positions along a dielectric waveguide, pinching antennas offer unprecedented channel reconfigurability and the ability to provide line-of-sight (LoS) links in scenarios with severe LoS blockages. The performance of pinching-antenna systems is highly dependent on the optimized placement of the pinching antennas, which must be jointly considered with traditional resource allocation (RA) variables---including transmission power, time slots, and subcarriers. The resulting joint RA problems are typically non-convex with complex variable coupling, necessitating sophisticated optimization techniques. This article provides a comprehensive survey of existing RA algorithms designed for pinching-antenna systems, supported by numerical case studies that demonstrate their potential performance gains. Key challenges and open research problems are also identified to guide future developments in this emerging field.
\end{abstract}

\IEEEpeerreviewmaketitle

\section{Introduction} \label{Sec:Introduction}
Pinching antennas have recently received significant attention as a promising technology for reconfiguring wireless propagation environments \cite{ding2024}. In pinching-antenna systems, an antenna can be dynamically activated on a dielectric waveguide tailored to the served user's location, enabling the establishment of strong line-of-sight (LoS) communication \cite{ding2024}. This approach offers greater channel reconfigurability compared to conventional movable or fluid antenna systems, which are typically limited to displacements on the order of a few wavelengths \cite{ding2024}. Pinching-antenna systems are also featured by their spatial flexibility, where  the configuration of a pinching-antenna array can be flexibly changed, e.g., adding or removing one antenna element is straightforward \cite{ding2024}.

The application of pinching antennas in high-frequency communication systems, such as those operating in the millimeter-wave (mmWave) and terahertz (THz) bands, is particularly compelling. These frequency bands offer wide bandwidths but suffer from severe propagation limitations, including high path loss and poor diffraction. In other words, mmWave and THz links are highly vulnerable to LoS blockages. By strategically deploying pinching antennas, it is possible to dynamically (re)-establish LoS paths, thereby enhancing system performance and mitigating communication outages.

The performance of pinching-antenna systems is highly sensitive to the spatial placement of the pinching antennas. As such, optimizing their locations constitutes a critical component in the system design process. Furthermore, this spatial optimization must be performed jointly with the allocation of conventional wireless communication resources, such as transmission power, time slots, and subcarriers. This collective process is referred to as resource allocation (RA) and forms the central focus of this article.
In general, RA problems in pinching-antenna systems are more challenging than those in conventional wireless networks, since they are non-convex optimization problems involving tightly coupled variables. Consequently, advanced mathematical transformations or approximation techniques are often required to derive high-quality, tractable solutions.

To date, the majority of the research efforts on pinching-antenna systems have concentrated on RA, given its pivotal role in unlocking the full potential of this emerging technology. In this context, this article provides a comprehensive review of the current landscape of RA strategies in pinching-antenna systems and identifies key challenges and future research directions. Specifically, it:
\begin{itemize}
    \item Presents a categorized survey of existing RA algorithms tailored for pinching-antenna systems;
    \item Provides informative numerical case studies to illustrate the performance gains enabled by pinching antennas for various user cases;
    \item Identifies open problems and highlights critical challenges that need to be addressed in future work.
\end{itemize}

\begin{figure}[ht!]
\centering
\includegraphics[width=0.95\linewidth]{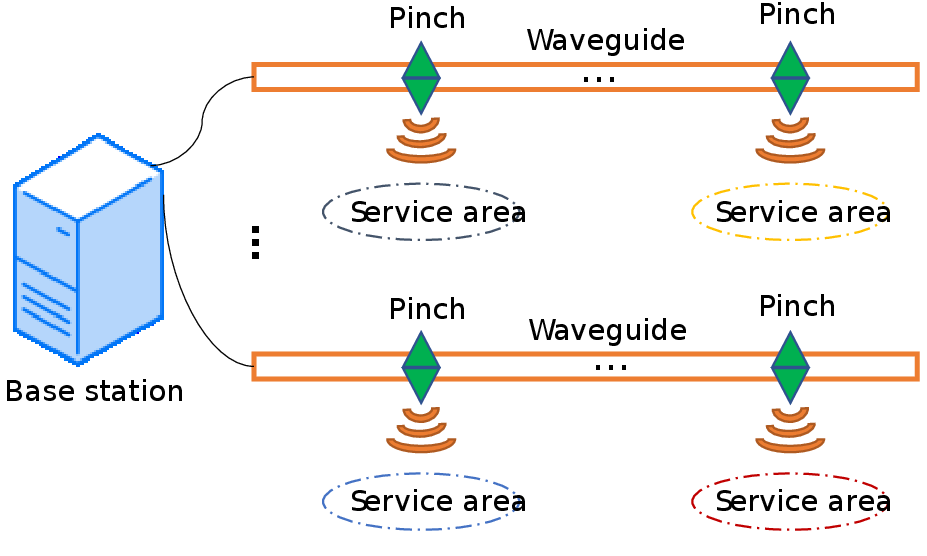}
\caption{Pinching-antenna system.} 
\label{fig_system}
\end{figure}

\section{Fundamentals of Pinching-Antenna Systems}
\label{Sec:LC}
First introduced by NTT Docomo in 2022, pinching-antenna systems have emerged as a new candidate for flexible-antenna systems. Unlike conventional fluid or movable antenna systems---whose spatial reconfigurability is constrained to the wavelength scale---the pinching antenna enables arbitrary positioning along a dielectric waveguide, thereby supporting large-scale reconfiguration of the antenna's location. As illustrated in Fig.~\ref{fig_system}, a typical pinching-antenna system operates by feeding user signals into a dielectric waveguide and radiating them through a leave-wave-like antenna formed by physically pinching the waveguide with a dielectric element. By strategically positioning the pinch-induced radiating elements, a strong LoS communication link can be dynamically established or restored, enhancing the overall performance of the wireless system. This technique is particularly advantageous in high-frequency bands, where communication links are more susceptible to LoS blockages due to pronounced signal attenuation and limited diffraction capability. Moreover, beyond enabling flexible antenna positioning, the system allows for flexible reconfiguration of the number of antennas by simply attaching or removing pinching elements along the waveguide. This inherent flexibility offers a novel, low-cost, scalable approach to implementing multi-input multi-output (MIMO) architectures and facilitates pinching-based beamforming---a novel technique that enhances communication performance by dynamically optimizing antenna locations for improved spatial directivity and signal quality. 

Since NTT Docomo’s initial demonstration in 2022—where the pinching-antenna system was successfully used to enable video transmission over a 60 GHz channel—research in this area has gained significant momentum, encompassing various aspects, such as physics-based modeling of the pinching antenna \cite{wang2025modeling}, channel estimation, performance analysis, RA, and the integration with other technologies. Among these, RA plays a pivotal role in realizing the full potential of pinching-antenna systems.
In contrast to conventional wireless systems, which primarily optimize transmission power, subcarriers, and time slots, pinching-antenna systems introduce an additional optimization degree of freedom: the position of the pinching antennas. Optimizing this spatial configuration is often interdependent with other RA variables, necessitating joint optimization strategies that can handle these complex relationships. The optimal allocation strategy depends heavily on the specific system configuration.
In the following, we consider three representative scenarios: (i) a single pinching antenna, (ii) multiple pinching antennas along a single waveguide, and (iii) multiple waveguides. Each case highlights different trade-offs and design considerations for resource optimization.

\section{RA for Single Pinching-Antenna Systems}
\label{RA_single PA}
In pinching-antenna systems, it is commonly assumed that user devices are equipped with a single antenna. When the base station (BS) also employs a single pinching antenna, the system reduces to a single-input single-output (SISO) configuration, where the transmitted signal is a scalar rather than a vector—thus simplifying analytical treatments and optimization.
Users are typically modeled as being randomly distributed within a rectangular service region, where LoS links can be established using the pinching antenna. Given that non-LoS components are significantly attenuated in high-frequency bands, they are often neglected in channel modeling. Therefore, by considering the LoS component only, the near-field free-space path loss model can be used to compute the channel gain.

%In pinching-antenna systems, it is often assumed that the users are equipped with single antenna. When the BS is equipped with a single pinching antenna, we have the so-called single-input single-output (SISO) system, where the signal is in the form of scalar rather than vector, simplifying the analysis. Moreover, it is often assumed that the users are randomly distributed in a rectangular region, where LoS links can be established using the pinching antenna. By ignoring the non-LoS (NLoS) path components, whose signal strengths are much weaker in high-frequency bands, the channel gain can be calculated using the near-field free space pathloss model, considering only the LoS component. 

\subsection{State-of-the-art}
In the simplest case with a single user and negligible waveguide propagation loss, it can be shown that the optimal pinching-antenna location is directly aligned with that of the user. This placement minimizes the transmission distance and, consequently, maximizes the channel gain \cite{tyrovolas2025}. However, when waveguide propagation loss is considered, direct alignment may no longer yield the optimal configuration. In such cases, the optimal antenna position can be determined by computing the first-order derivative of the channel gain with respect to the antenna location \cite{tyrovolas2025}. Nonetheless, as noted in \cite{tyrovolas2025}, direct alignment remains a near-optimal strategy due to the low attenuation coefficient of dielectric waveguides. 

In multi-user pinching-antenna systems, multiple access schemes, such as time division multiple access (TDMA) and non-orthogonal multiple access (NOMA), should be employed to accommodate multiple users, due to the system's constraint of having only a single antenna. For TDMA, two configurations are typically considered: 1) The pinching antenna location is dynamically adjusted for each user in their respective time slots; 2)
All users share a fixed pinching antenna location throughout the scheduling interval. The first configuration effectively reduces the system to a sequence of single-user scenarios, where the optimal antenna placement is to directly align the antenna with each individual user \cite{ding2024}. In contrast, the second configuration reduces the implementation complexity for frequently adjusting the antenna position, but introduces a more complex joint optimization problem, as a single antenna location must serve all users. This challenge is similar to the scenario in NOMA, where the antenna serves multiple users simultaneously.

In \cite{Zeng_COMML25}, the authors investigated the sum-rate maximization for a multi-user uplink system utilizing a single pinching antenna. They demonstrated that the maximum transmission power should be used by each user to achieve the highest sum rate. For both NOMA and TDMA with a common antenna location, the resulting sum-rate expression was shown to be a non-convex function, represented as a sum of generalized bell-shaped membership functions, which leads to the presence of multiple local optima. To solve this, one-dimensional search and the particle swarm optimization (PSO) algorithm were employed to obtain optimal and near-optimal antenna positions, respectively.
To address the tradeoff between energy consumption and system throughput, the authors in \cite{zeng2025EE} formulated an energy efficiency (EE) maximization problem for the same system. In this case, transmitting at full power is no longer optimal. Instead, the optimal transmit power for each user was computed using the Dinkelbach algorithm, assuming a fixed antenna position. Conversely, for a given power allocation, the optimal (near-optimal) antenna location was determined via a one-dimensional search (PSO). These two steps were iteratively carried out until convergence, yielding a jointly optimized power and antenna configuration.

\subsection{Numerical Case Study}
\begin{figure}[ht!]
\centering
\includegraphics[width=1\linewidth]{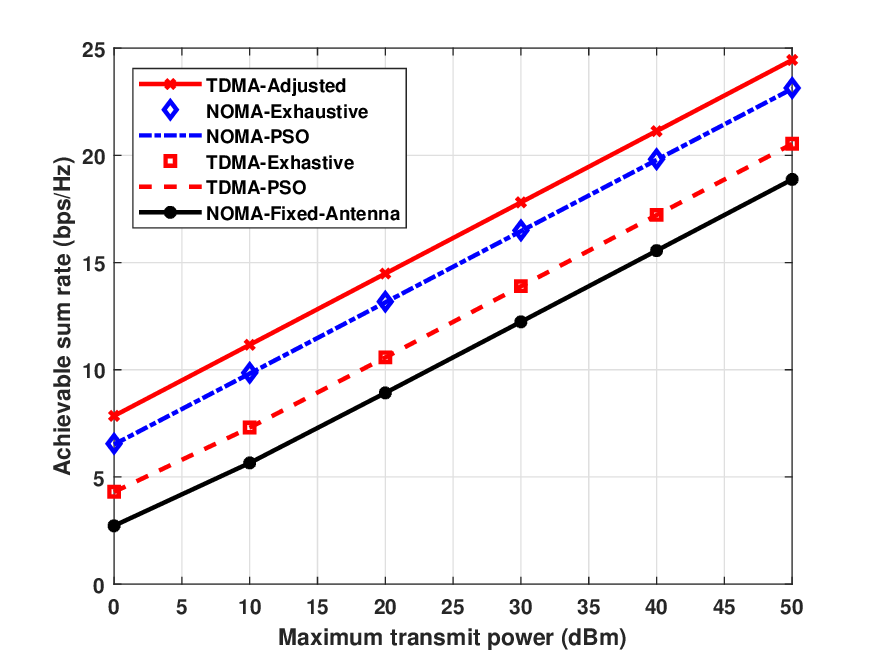}
\caption{Sum rate versus the maximum transmit power for an uplink multi-user system \cite{Zeng_COMML25}.} 
\label{fig_sum_rate}
\end{figure}

This subsection presents simulation results highlighting the performance improvements offered by pinching-antenna systems compared to traditional fixed-antenna setups. Figure~\ref{fig_sum_rate} illustrates the relationship between the achievable sum rate and the maximum transmit power per user in an uplink configuration with five users. 
Six transmission schemes are evaluated. In TDMA-Adjusted, the antenna is dynamically positioned for each user individually, with optimal alignment. In contrast, TDMA-Exhaustive and TDMA-PSO assume a common antenna position shared by all users, determined via exhaustive search and PSO, respectively. Similarly, NOMA-Exhaustive and NOMA-PSO apply these optimization methods within a NOMA framework. The NOMA-Fixed-Antenna scheme represents the conventional baseline with a static antenna placement.

The simulation results indicate that the pinching-antenna system significantly outperforms the conventional fixed-antenna baseline, primarily due to its ability to dynamically reposition the antenna closer to the users. Among the shared-position configurations, the NOMA-based schemes outperform their TDMA counterparts, reflecting the spectral efficiency gains of NOMA. However, the TDMA-Adjusted scheme achieves the highest performance, highlighting the benefits of user-specific antenna positioning. Furthermore, the PSO-based strategies yield results that closely match those obtained via exhaustive search, validating PSO as an efficient and effective heuristic for optimizing antenna placement in both NOMA and TDMA scenarios.

\section{RA for A Single Waveguide with Multiple Pinching Antennas} \label{RA_multi_PA}
%When the system is equipped with a single waveguide and multiple pinching antennas, the pinching antennas act as a linear antenna array. The signal radiated into the air by each pinching antenna is a phase-shifted version of the original one fed into the waveguide. The total field of the pinching antennas is then the superposition of the fields radiated by each antenna. To provide directive pattern, it is necessary that the partial fields (generated by the individual pinching antennas) interfere constructively in the desired direction and interfere destructively in the remaining space. Note that the linear array formed by the pinching antennas is not necessarily of uniform amplitude and spacing. Clearly, the spacing of linear array can be easily adjusted by moving the pinching antennas along the waveguide. In contrast, the power (amplitude) of each pinching antenna depends mostly on its length \cite{wang2025modeling}. By using coupled-mode theory, two power (amplitude) models are proposed for pinching antennas in \cite{wang2025modeling},: equal power and proportional power models. The former is achieved when each pinching antenna is of different length, and vice versa. 
When the system incorporates a single waveguide with multiple pinching antennas, the antennas collectively function as a linear antenna array. Each pinching antenna radiates a phase-shifted replica of the original signal injected into the waveguide. The resulting total radiated field is the superposition of the individual fields emitted by each antenna element. To achieve a directive radiation pattern, the fields generated by the individual pinching antennas must interfere constructively in the desired directions while undergoing destructive interference in all other directions.

It is important to note that the linear array formed by the pinching antennas does not necessarily exhibit uniform amplitude or spacing. The inter-element spacing can be readily adjusted by repositioning the antennas along the waveguide. However, the amplitude (or power) radiated by each pinching antenna primarily depends on its physical length, as described in \cite{wang2025modeling}. 
Based on the coupled-mode theory, two power (or amplitude) distribution models for pinching antennas were introduced in \cite{wang2025modeling}, namely the equal power model and the proportional power model. In the equal power model, antenna lengths are varied to ensure uniform power distribution among elements. Conversely, in the proportional power model, the use of antennas of identical length results in varying power outputs.

\subsection{State-of-the-art}
In \cite{Xu_WCL25}, the downlink rate maximization problem was investigated for a simplified scenario involving a single user, under the assumption of equal power distribution across all pinching antennas. In the absence of inter-user interference, the objective of maximizing the data rate reduces to maximizing the signal-to-noise ratio (SNR). To this end, it is optimal to utilize the maximum available transmit power. A two-stage algorithm was proposed for the placement of pinching antennas: the first stage focuses on minimizing large-scale path loss by optimizing antenna positions, while the second stage refines the locations to achieve constructive interference and mitigate mutual coupling effects. Furthermore, the authors of \cite{ding2024} demonstrated that under this placement strategy, the array gain of the pinching-antenna system scales linearly with the number of antennas.

In \cite{mu2025_multicast}, the pinching-antenna system was extended to a multicast communication scenario, where multiple pinching antennas along a single waveguide jointly transmit a common signal to multiple users. The design objective was to maximize the minimum SNR across all users for a given transmit power. Despite the absence of inter-user interference, the single-user optimization method proposed in \cite{Xu_WCL25} is not directly applicable, as the antenna configuration must simultaneously satisfy the reception requirements of multiple users. The resulting non-convex optimization problem was addressed using a heuristic PSO algorithm.

Beyond multicast scenarios, conventional multi-user communication using pinching antennas has been studied in \cite{ding2024, xie2025graphneural, Tegos_2025, wang2024}. Due to the shared single waveguide structure, multiple access techniques are needed to support multi-user transmission. For example, TDMA was adopted in \cite{ding2024, xie2025graphneural, Tegos_2025}, while NOMA was employed in \cite{wang2024}.

%Beyond multicast scenarios, conventional multi-user communication using pinching antennas has been studied in \cite{ding2024, xie2025graphneural, Tegos_2025, wang2024}. Due to the shared single waveguide structure, MIMO multiplexing is not possible, necessitating the use of multiple access techniques for supporting multi-user transmission. TDMA was adopted in \cite{ding2024, xie2025graphneural, Tegos_2025}, while NOMA was employed in \cite{wang2024}.

In TDMA-based systems, two distinct scenarios are considered: (1) the pinching antennas are reconfigurable for each individual user, and (2) all users share a common, fixed antenna placement. In the first scenario, the multi-user system can be decomposed into multiple independent single-user systems, allowing the direct application of the optimal antenna placement strategy from \cite{Xu_WCL25} to maximize each user's channel gain \cite{ding2024}. In the second scenario, although inter-user interference is absent, the joint optimization of a single antenna configuration to serve multiple users becomes significantly more challenging. To address this, the authors of \cite{xie2025graphneural} formulated a downlink EE maximization problem and modeled the system as a bipartite graph. A graph attention network-based framework was proposed to jointly optimize antenna placement and power allocation.

For the uplink scenario, the authors of \cite{Tegos_2025} addressed the problem of maximizing the minimum user data rate to balance the trade-off between system throughput and user fairness. Due to the coupling of variables in the formulated optimization problem, a block coordinate descent (BCD) framework was adopted to decouple the problem into two subproblems: one for optimizing antenna placement and the other one for power allocation. The former was solved iteratively by successive convex approximation (SCA), while a semi-analytical solution was derived for the latter.

Additionally, the authors of \cite{wang2024} considered a NOMA-assisted system for downlink communication, with the objective of maximizing the system sum rate. In this setup, multiple pinching antennas can be selectively activated at pre-configured locations along the waveguide. The joint optimization of the number and positions of active antennas was formulated as a one-sided one-to-one matching problem, and a low-complexity, matching-based antenna activation algorithm was developed to solve it efficiently.

\subsection{Numerical Case Study}
This subsection presents numerical results to demonstrate the performance benefits of deploying multiple pinching antennas. We consider a downlink multi-user scenario, with the objective of maximizing the system's EE, subject to the users' individual rate requirement. Following the approach in \cite{ding2024}, TDMA with reconfigurable pinching antenna locations per user is adopted to enhance the channel gains. However, unlike \cite{ding2024, xie2025graphneural}, the time allocation for each user is not assumed to be uniform but instead treated as an optimization variable. Consequently, a joint optimization problem is formulated with respect to the users’ transmission times, power allocations, and pinching antenna positions.
To address the non-convex nature of the problem, we first determine the optimal antenna positions using the strategy proposed in \cite{Xu_WCL25}. Based on these positions, the remaining problem can be further decomposed into two subproblems---power allocation and time allocation---which are solved iteratively until convergence. The power allocation subproblem is solved via an iterative method that yields a semi-analytical solution in each iteration. Similarly, a semi-analytical approach is employed to solve the time allocation subproblem.

\begin{figure*}[t!]
    \centering
    \begin{subfigure}[t]{0.48\textwidth}
        \centering
        \includegraphics[width=1\textwidth]{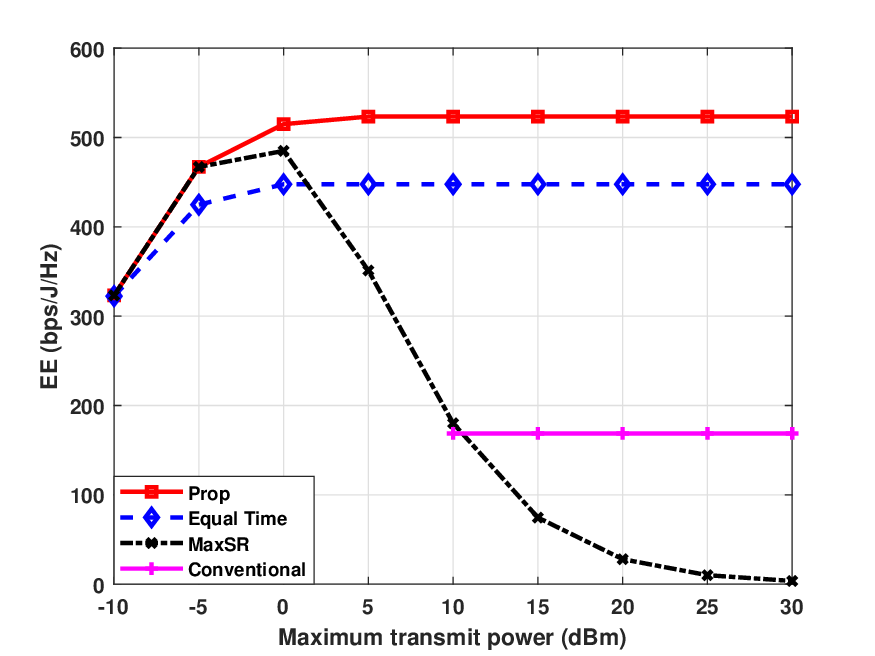}
        \caption{}
        \label{fig:Low_2}
    \end{subfigure}%
    ~ 
    \begin{subfigure}[t]{0.55\textwidth}
        \centering
        \includegraphics[width=1\textwidth]{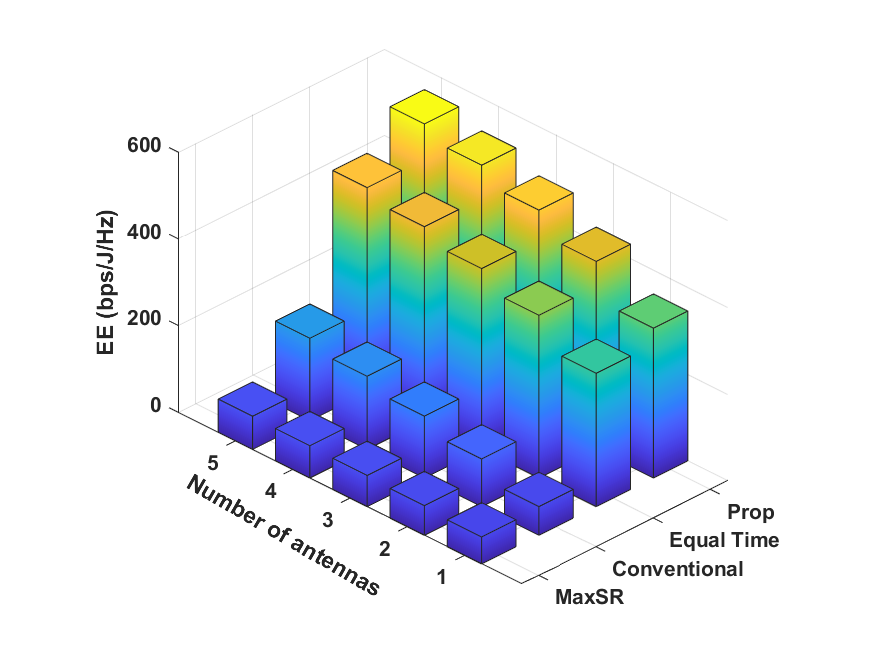}
        \caption{}
        \label{fig:Low_3}
    \end{subfigure}
    \caption {EE versus a) maximum transmit power constraint at the BS; and b) the number of pinching antennas.}
    \label{fig:EE}
\end{figure*}

Figure \ref{fig:EE}(a) reveals how the system's EE varies with the maximum transmit power constraint $P_{\max}$
for the different schemes under evaluation. Throughout the considered range of $P_{\max}$, all configurations utilizing pinching antennas are able to consistently meet the users’ quality-of-service (QoS) requirements. In contrast, the traditional fixed-antenna approach fails to satisfy these constraints when $P_{\max}$ is below 10 dBm, underscoring the superior performance of the pinching-antenna architecture. As 
$P_{\max}$ increases, the EE of both the proposed method (``Prop'') and the equal time allocation approach (``Equal Time'') improves initially and then stabilizes. On the other hand, the sum rate maximization scheme (``MaxSR'') shows a peak in EE before declining at higher power levels. These observations suggest that prioritizing EE leads to a more balanced trade-off between data rate and power usage compared to schemes focused solely on throughput. Moreover, across all 
$P_{\max}$ values, the ``Prop'' approach consistently yields higher EE than the ``Equal Time'' scheme, mainly due to the benefits of optimized time allocation across users.

Figure \ref{fig:EE}(b) explores how the system’s EE evolves with the number of pinching antennas. As expected, increasing the number of antennas leads to improved EE for all examined strategies. Among them, the proposed scheme achieves the highest EE, followed by ``Equal Time''. 
In contrast, the EE improvement of the ``MaxSR'' approach is significantly slower than that of the three EE-maximization-based schemes, further highlighting the importance of explicitly optimizing for EE.
%The increment in EE for ``MaxSR'' is much slower than that of the three schemes based on EE maximization, again showing the necessity of EE maximization. 

\section{RA for Multiple Waveguides}
\label{RA_multi_Waveguide}
When the BS is equipped with multiple waveguides, each connected to a dedicated radio frequency (RF) chain, MIMO multiplexing can be employed to enable multi-user data transmission. Depending on the number of pinching antennas integrated within each waveguide, two scenarios arise: 1) a single pinching antenna per waveguide, and 2) multiple pinching antennas per waveguide. In the first case, conventional MIMO multiplexing is sufficient. In the second case, however, MIMO multiplexing must be jointly designed with pinching beamforming to effectively support multi-user transmission. Furthermore, the signal fed into each waveguide may either represent an independent data stream for a single user or a superposition of multiple data streams intended for different users.

\subsection{State-of-the-art}
In \cite{ding2024}, the authors investigated a downlink system comprising multiple waveguides, each equipped with a single pinching antenna. For the special case of two users, they first established an upper bound on the achievable rate, assuming that each user fully occupies the available bandwidth without causing interference to the other user. Based on this, two necessary conditions for achieving the upper bound were derived: the phase-matching condition and the orthogonality condition. Furthermore, specific design examples were provided in which the upper bound is attained through the joint optimization of the pinching antenna locations and the digital beamforming vectors. These results clearly demonstrate the advantages of pinching-antenna architectures over conventional fixed-antenna systems, where completely avoiding co-channel interference and maximizing a user's effective channel gain cannot be accomplished simultaneously.%inter-user interference is generally non-negligible and more difficult to suppress.

In \cite{hu2025}, the authors addressed the problem of sum-rate maximization in a multi-user multiple-input single-output (MISO) NOMA system employing multiple dielectric waveguides. To more accurately model the system, in-waveguide propagation loss was explicitly considered in the channel gain characterization. The resulting optimization problem is non-convex and involves coupled variables. To tackle this challenge, the alternating optimization (AO) algorithm was used to decompose the original problem into two subproblems: one for power allocation and the other one for optimizing the pinching antenna positions. The SCA technique was then applied to obtain suboptimal solutions for both subproblems. Simulation results show that the proposed pinching-antenna system not only outperforms the conventional fixed-antenna architecture but also yields superior performance compared to a naive pinching-antenna design that neglects waveguide power attenuation.

%In the above works \cite{ding2024} and \cite{hu2025}, it is assumed that each waveguide is equipped with a single pinching antenna. The more general case where each waveguide is equipped with multiple pinching antennas is considered in \cite{wang2025modeling, zhao2025waveguide, bereyhi2025, sun2025PLS, xu2025joint }. More exactly, in \cite{zhao2025waveguide}, the authors consider a multi-user scenario, in which each user is served by one dedicated allocated waveguide. The objective is maximize the system's sum rate, by jointly optimizing the power allocation to the users and placement of the pinching antennas. Two pinching antenna activations are considered, one with continuous activation, while the other with discrete activation. Similar to \cite{hu2025}, the AO algorithm is employed to decompose the original problem into two subproblems. For the power allocation subproblem, the SCA technique is invoked. For the pinching beamforming design subproblem, a penalty-based gradient ascent algorithm is developed for the continuous activation case, while a matching theory-based algorithm is proposed for the discrete activation case. 

In the aforementioned works \cite{ding2024, hu2025}, it was assumed that each waveguide is equipped with a single pinching antenna. The more general case, where each waveguide is equipped with multiple pinching antennas, has been explored in \cite{wang2025modeling, zhao2025waveguide, bereyhi2025, sun2025PLS, xu2025joint}. Specifically, \cite{zhao2025waveguide} investigated a multi-user scenario in which each user was served via a dedicated waveguide. The primary objective was to maximize the system sum rate through a joint optimization of user power allocation and the spatial placement of pinching antennas. Two activation strategies for the pinching antennas were considered: continuous activation and discrete activation. 
To address the resulting non-convex optimization problem, an AO framework was adopted, similar to the approach in \cite{hu2025}. The power allocation subproblem was addressed using the SCA method. For the pinching beamforming design subproblem, different solution strategies were employed based on the activation type: a penalty-based gradient ascent algorithm was developed for the continuous activation case, while a matching theory-based algorithm was proposed for the discrete activation scenario.

In contrast to \cite{zhao2025waveguide}, which did not involve digital beamforming, the works in \cite{wang2025modeling, bereyhi2025, sun2025PLS, xu2025joint} adopted a hybrid beamforming architecture that integrates digital beamforming at the BS with pinching-antenna assisted analog beamforming realized through the spatial configuration of pinching antennas. Among these, \cite{bereyhi2025} addressed the weighted sum-rate maximization problem for both downlink and uplink transmission scenarios.
For the downlink, digital beamforming was optimized using fractional programming, while the positions of the pinching antennas were optimized via an iterative one-dimensional search based on the Gauss-Seidel method. The digital and analog beamforming components were updated alternately until convergence. For the uplink, a similar solution framework was applied, where the optimal digital combiner at the BS is based on the well-known minimum mean squared error (MMSE) detector.

In \cite{sun2025PLS}, the authors focused on maximizing the weighted secrecy sum-rate in a pinching-antenna system with multiple legitimate users and multiple eavesdroppers. To solve the resulting non-convex problem, a fractional programming-based BCD algorithm was proposed. Within this framework, the digital beamformer was optimized using the Lagrange multiplier method, while the pinching antenna locations were determined via an iterative one-dimensional search, following the approach used in \cite{bereyhi2025}.

Additionally, the authors in \cite{xu2025joint} formulated a sum-rate maximization problem for a downlink multi-user MISO system and explored both optimization-based and learning-based methodologies. For the optimization-based approach, a majorization-minimization and penalty dual decomposition (MM-PDD) algorithm was developed. This method addresses the non-convexity arising from the complex exponential terms by introducing a Lipschitz-continuous surrogate function, followed by problem decoupling using the PDD framework. For the learning-based approach, a novel Karush-Kuhn-Tucker (KKT)-guided dual learning method was proposed. This approach enables data-driven reconstruction of KKT-optimal solutions by learning the dual variables. Simulation results demonstrated that the learning-based approach achieves superior performance compared to the MM-PDD optimization counterpart, highlighting the potential of integrating model-based and data-driven techniques for pinching-antenna system designs. 

In contrast to the aforementioned works \cite{zhao2025waveguide, bereyhi2025, sun2025PLS, xu2025joint}, which primarily addressed sum-rate maximization, the study in \cite{wang2025modeling} focused on transmit power minimization under both equal power and proportional power models. 
To solve this problem, the authors proposed a penalty-based AO algorithm. Within this framework, the positions of the pinching antennas were optimized through an iterative one-dimensional search, following the method adopted in \cite{bereyhi2025}.  Simulation results demonstrated that pinching-antenna systems could achieve substantial transmit power savings compared to conventional and massive MIMO systems. Moreover, the performance under the proportional power model was shown to be comparable to that under the equal power model.

\begin{figure*}[ht!]
\centering
\includegraphics[width=1\linewidth]{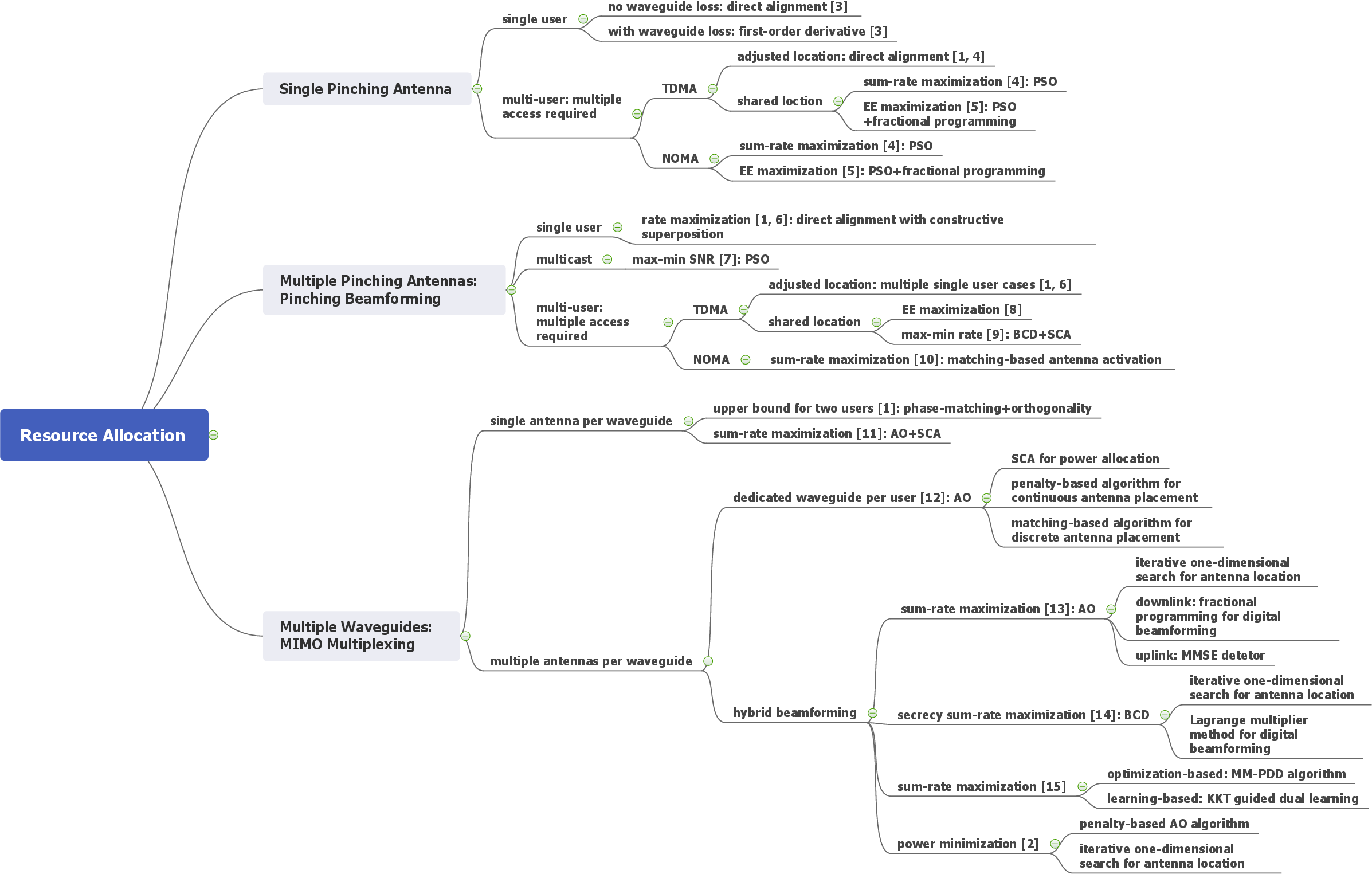}
\caption{Summary of the existing RA schemes for pinching-antenna systems.} 
\label{fig_summary}
\end{figure*}

A summary of the existing RA schemes for pinching-antenna systems is provided in Fig. \ref{fig_summary}.

\subsection{Numerical Case Study}
In this subsection, we present a numerical case study to demonstrate the performance advantages of pinching-antenna systems over conventional fixed-antenna configurations in a multi-waveguide setting. We consider an uplink scenario involving three users, served by three pinching antennas, each activated on a different waveguide. The system design objective is to maximize the sum rate subject to a per-user maximum transmit power constraint.

Given fixed antenna locations, the MMSE detector is known to be optimal \cite{bereyhi2025} and is therefore employed for both pinching-antenna and conventional fixed-antenna systems. For optimizing the pinching antenna locations, we evaluate three schemes that represent different trade-offs between performance and computational complexity.

\begin{figure}[ht!]
\centering
\includegraphics[width=1\linewidth]{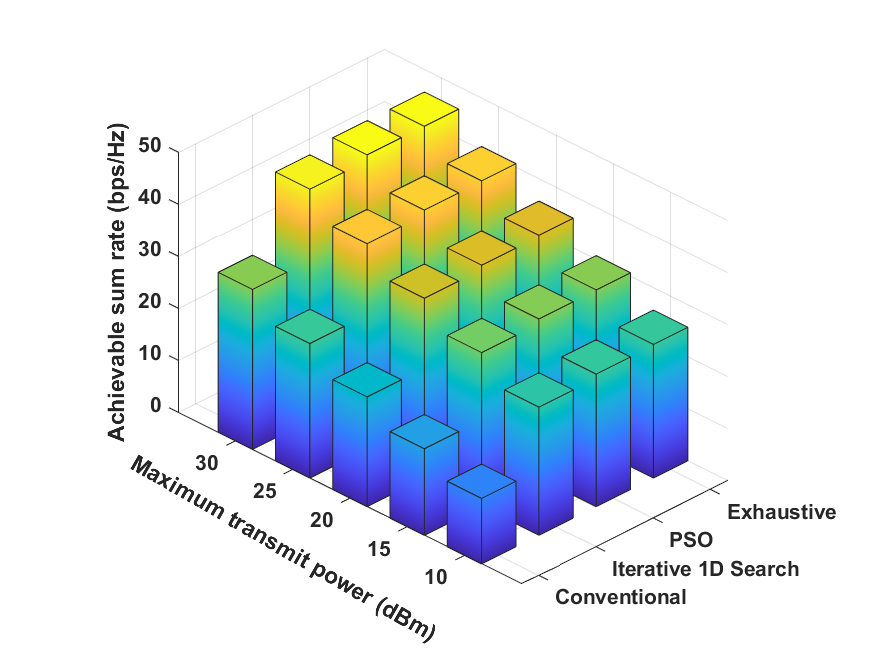}
\caption{Achievable sum rate versus the maximum transmit power at the users.} 
\label{fig_multi}
\end{figure}

As illustrated in Fig.~\ref{fig_multi}, the highest sum rate is achieved through an exhaustive search over all possible antenna positions, albeit at the cost of significant computational overhead. The PSO algorithm closely approximates the performance of the exhaustive search, demonstrating its near-optimality with reduced complexity. In contrast, the iterative one-dimensional search method, as proposed in \cite{sun2025PLS, wang2025modeling, bereyhi2025}, incurs a performance gap relative to exhaustive search but offers greater computational efficiency.

Importantly, all three pinching-antenna-based schemes significantly outperform the conventional fixed-antenna system, thereby reaffirming the effectiveness and potential of the pinching-antenna approach in multi-waveguide scenarios.

%In this subsection, numerical case study is conducted to show the superiority of pinching-antenna systems over conventional fixed-antenna systems in the case of multi-waveguide. We consider an uplink scenario with three users, which are served by three pinching antennas, each activated on a separate waveguide. The objective is to maximize the system's sum rate, under given maximum transmit power constraint at the users.Under given antenna locations, the MMSE detector is optimal \cite{bereyhi2025}, and thus, is adopted for both pinching-antenna systems and conventional fixed-antenna system. For the pinching antenna location optimization, three schemes are considered, highlighting different tradeoff of performance and complexity. As show in Fig.~\ref{fig_multi}, the highest sum rate is obtained by the exhaustive search over all antenna positions, which is also the most computationally complex. The heuristic PSO algorithm almost overlaps with exhaustive search, showing its near-optimality. The iterative one-dimensional search adopted by \cite{sun2025PLS, wang2025modeling,  bereyhi2025} exhibits a performance gap with the exhaustive search, but is more computationally efficient. Finally, the three pinching-antenna-based schemes significantly outperforms its conventional fixed-antenna counterpart, again validating the effectiveness of pinching antennas. 

\section{Challenges and Open Directions}
Despite the substantial progress made in the design and optimization of pinching-antenna systems, RA remains an emerging and underexplored research area. The unique characteristics of these systems---such as spatial flexibility of antennas and tight coupling among system parameters---introduce new complexities that are not present in conventional fixed-antenna architectures. The following four key research directions present both challenges and opportunities for advancing pinching-antenna systems. 

\subsection{RA for Orthogonal Frequency Division Multiplexing (OFDM)-based Systems}
Orthogonal frequency division multiplexing (OFDM) is a foundational technology in modern wireless communication systems. Given that pinching antennas are most likely to be deployed in high-frequency bands---such as mmWave and THz---the operating bandwidth tends to be wide, encompassing a large number of subcarriers. However, most existing works assume a flat frequency response across the entire bandwidth as the signal propagates through the waveguide. This assumption may not hold in practical scenarios. As highlighted in \cite{hu2025}, power attenuation in dielectric waveguides is frequency-dependent, suggesting that frequency selectivity should be taken into account. To accurately capture this characteristic, further research is needed in channel modeling. Moreover, to effectively manage the frequency-selective nature of the waveguide, subcarrier allocation should be integrated into the existing RA frameworks to enhance system performance and robustness.

\subsection{Robust RA under Imperfect User Locations}
In most existing studies, it is assumed that the coordinates of all serving users are perfectly known at the BS. While this assumption simplifies the RA design---particularly for systems relying on pinching antenna positioning and directional beamforming---it is often unrealistic in practical scenarios. In real-world deployments, user location information is typically obtained through positioning or tracking systems, which are inherently subject to measurement errors, estimation delays, and environmental uncertainties. Such inaccuracies in user location can significantly degrade system performance, especially in high-frequency bands (e.g., mmWave and THz), where beam misalignment due to localization errors can lead to severe signal attenuation and co-channel interference.

Therefore, there is a pressing need for further research into robust RA strategies that explicitly account for uncertain or imperfect user location information. This includes developing optimization frameworks that incorporate location uncertainty models, designing robust beamforming algorithms that are resilient to positioning errors, and investigating trade-offs between localization accuracy and communication performance. Integrating such robustness into the system design will be critical for ensuring reliable and efficient operation of pinching-antenna systems in dynamic and imperfect real-world environments.

\subsection{Machine Learning-Empowered RA}
RA in pinching-antenna systems is inherently challenging due to the strong coupling among variables—such as transmit power, beamforming vectors, and antenna positions—and the highly non-convex nature of the associated optimization problems. These often involve complex constraints and nonlinear interactions, making them difficult to solve using standard convex techniques. To address this, researchers commonly adopt advanced approximation methods, such as SCA, fractional programming, or penalty-based decomposition to obtain tractable formulations. Yet, optimizing pinching antenna positions remains difficult, involving combinatorial complexity (in discrete activation) or high-dimensional search spaces (in continuous activation), often leading to high computational cost and suboptimal solutions.

In this context, machine learning (ML) offers a promising alternative. ML methods---especially deep learning, and KKT-guided learning---can learn mappings or optimization policies directly from data, reducing reliance on iterative procedures. Preliminary results (e.g., \cite{xu2025joint}) show that ML-based approaches can outperform conventional methods in performance. However, research on ML-driven RA for pinching-antenna systems is still in its infancy. Future efforts should focus on designing generalizable learning architectures, hybrid model- and data-driven frameworks, and improving robustness and interpretability. These advancements are essential for realizing scalable and efficient pinching-antenna systems in next-generation wireless networks.

{\color{black}
\subsection{RA for Pinching-Antenna Assisted Integrated Sensing and Communication}
Integrated sensing and communication (ISAC) is emerging as a critical paradigm for next-generation wireless networks, by uniting wireless communication and sensing functionalities into a single system. Pinching antenna systems significantly benefit ISAC through dynamic reconfigurability along dielectric waveguides, allowing optimized LoS creation, enhanced spatial flexibility for improved sensing (e.g., target diversity), and cost-effective deployment. However, the unique capabilities of pinching antennas introduce complex challenges in RA for ISAC. The fundamental trade-off between maximizing communication performance (e.g., data rate) and sensing performance (e.g., accuracy) demands sophisticated optimization algorithms, often leading to high computational complexity. Moreover, dynamically acquiring accurate channel state information for both communication and sensing links, as well as managing interference arising from shared resources, are crucial practical considerations. Addressing these challenges effectively will be key to realizing the full potential of pinching-antenna-assisted ISAC systems.}

\section{Conclusion} 
\label{Sec:Conclusion}
This article has presented a categorized overview of RA algorithms for pinching-antenna systems, highlighting their pivotal role in fully realizing the performance gains offered by this emerging technology. Through numerical case studies, we have demonstrated that pinching-antenna systems consistently outperform their conventional fixed-antenna counterparts for various deployment scenarios, including configurations with a single pinching antenna, multiple antennas along a single waveguide, and multiple waveguides. Furthermore, several promising research directions have been identified, including RA strategies for multi-carrier systems, robust RA under user location uncertainty, ML-empowered RA frameworks and RA for pinching-antenna-assisted ISAC systems. These avenues hold significant potential for advancing the capabilities of pinching-antenna systems in future wireless networks.

%This article provided an overview of the RA algorithms for pinching-antenna systems in a categorized fashion. It was clear that the RA algorithms play a pivotal role in achieving the maximum benefits of pinching antennas. Provided numerical case studies demonstrated that pinching-antenna systems could obtain superior performance over existing fixed-antenna counterpart in scenarios with a single pinching antenna, multiple pinching antennas over a single waveguide and multiple waveguides. Finally, RA in OFDM-based multi-carrier systems, robust RA under inaccurate user locations and ML-empowered RA are identified and discussed as potential research directions on the RA for pinching-antenna systems. 

\bibliographystyle{IEEEtran}
\bibliography{biblio}

\balance

\end{document}